%% file: ShoeLASIQ.tex
\documentclass[aip,apl,amsmath,amssymb,reprint]{revtex4-2}

\input{Preamble/packages}

\newcommand{\QuTech}{\affiliation{QuTech, Delft University of Technology, P.O. Box 5046, 2600 GA Delft, The Netherlands}}
\newcommand{\Kavli}{\affiliation{Kavli Institute of Nanoscience, Delft University of Technology, P.O. Box 5046, 2600 GA Delft, The Netherlands}}
\newcommand{\TNO}{\affiliation{Netherlands Organisation for Applied Scientific Research (TNO), P.O. Box 96864, 2509 JG The Hague, The Netherlands}}
\newcommand{\nametitle}{Post-fabrication frequency trimming of coplanar-waveguide resonators in circuit QED quantum processors}
\newcommand{\SSRO}{\mathrm{SSRO}}
\newcommand{\NbTiN}{\mathrm{NbTiN}}
\newcommand{\Silicon}{\mathrm{Si}}
\newcommand{\lambdaovfour}{\lambda/4}
\newcommand{\Strans}{S_{\mathrm{21}}}
\newcommand{\drivefreq}{f}
\newcommand{\freqinitial}{f_{\mathrm{0}}}
\newcommand{\fR}{f_{\mathrm{R}}}
\newcommand{\fP}{f_{\mathrm{P}}}
\newcommand{\fQ}{f_{\mathrm{Q}}}
\newcommand{\DeltaRQ}{\Delta_{\mathrm{RQ}}}
\newcommand{\DeltaR}{\Delta_{\mathrm{R}}}
\newcommand{\DeltaP}{\Delta_{\mathrm{P}}}
\newcommand{\DeltaPR}{\Delta_{\mathrm{PR}}}
\newcommand{\gammaP}{\gamma_{\mathrm{P}}}
\newcommand{\gammaR}{\gamma_{\mathrm{R}}}
\newcommand{\couplingJ}{J}
\newcommand{\kappaP}{\kappa}
\newcommand{\kappaeff}{\kappa_{\mathrm{eff}}}
\newcommand{\kappaDrive}{\kappa_{\textrm{drive}}}
\newcommand{\dispshift}{2\chi}
\newcommand{\dispshifteff}{2\chi_{\mathrm{eff}}}
\newcommand{\Deltaf}{\Delta f}
\newcommand{\Deltal}{\Delta l}
\newcommand{\Qc}{Q_{\mathrm{c}}}
\newcommand{\Qi}{Q_{\mathrm{i}}}
\newcommand{\tauro}{\tau_{\mathrm{RO}}}
\newcommand{\taup}{\tau_{\mathrm{p}}}
\newcommand{\tauwait}{\tau_{\mathrm{wait}}}
\newcommand{\vphase}{\nu_\rho}
\newcommand{\Ec}{E_{\mathrm{c}}}
\newcommand{\Ej}{E_{\mathrm{J}}}
\newcommand{\resJ}{R_{\mathrm{J}}}
\newcommand{\ncrit}{n_{\mathrm{crit}}}
\newcommand{\Fro}{F_{\mathrm{RO}}}
\newcommand{\epsro}{\epsilon_{\mathrm{RO}}}
\newcommand{\pqndpi}{P_{\mathrm{QND}, \pi}}
\newcommand{\mps}{\mathrm{m}/\mathrm{s}}
\newcommand{\um}{\upmu \mathrm{m}}
\newcommand{\us}{\upmu \mathrm{s}}
\newcommand{\GHz}{\mathrm{GHz}}
\newcommand{\MHz}{\mathrm{MHz}}
\newcommand{\ns}{\mathrm{ns}}
\newcommand{\percent}{\%}
\newcommand{\mW}{\mathrm{mW}}
\newcommand{\Ohm}{\Omega}
\newcommand{\Celsius}{^\circ\mathrm{C}}
\newcommand{\nm}{\mathrm{nm}}
\newcommand{\nA}{\mathrm{nA}}
\newcommand{\seconds}{\mathrm{s}}
\newcommand{\minutes}{\mathrm{min}}
\newcommand{\hours}{\mathrm{h}}
\newcommand{\Watt}{\mathrm{W}}
\newcommand{\uCcm}{\mathrm{\upmu C/cm^2}}
\newcommand{\rpm}{\mathrm{rpm}}

\begin{document}

\title{\nametitle}
\author{S.~Vall\'es-Sanclemente}\thanks{These authors contributed equally to this work.}\QuTech\Kavli
\author{S.~L.~M.~van~der~Meer}\thanks{These authors contributed equally to this work.}\QuTech\Kavli
\author{M.~Finkel}\QuTech\Kavli
\author{N.~Muthusubramanian}\QuTech\Kavli
\author{M.~Beekman}\QuTech\TNO
\author{H.~Ali}\QuTech\Kavli
\author{J.~F.~Marques}\QuTech\Kavli
\author{C.~Zachariadis}\thanks{Present address: Quantware B.V., Elektronicaweg 10, 2628 XG Delft, The Netherlands}\QuTech\Kavli
\author{H.~M.~Veen}\QuTech\Kavli
\author{T.~Stavenga}\QuTech\Kavli
\author{N.~Haider}\QuTech\TNO
\author{L.~DiCarlo}\QuTech\Kavli

\date{\today}

\begin{abstract}
We present the use of grounding airbridge arrays to trim the frequency of microwave coplanar-waveguide (CPW) resonators post fabrication. This method is compatible with the fabrication steps of conventional CPW airbridges and crossovers and increases device yield by allowing compensation of design and fabrication uncertainty with $100~\MHz$ range and $10~\MHz$ resolution. We showcase two applications in circuit QED. The first is elimination of frequency crowding between resonators intended to readout different transmons by frequency-division multiplexing. The second is frequency matching of readout and Purcell-filter resonator pairs. Combining this matching with transmon frequency trimming by laser annealing reliably achieves fast and high-fidelity readout across 17-transmon quantum processors.
\end{abstract}
\maketitle

\begin{bibunit}[apsrev4-2]

Accurate targeting of qubit and resonator frequencies is increasingly important as circuit QED processors scale.
Frequency errors arising from design limitations and fabrication uncertainly will otherwise bottleneck the yield of fully-operable devices~\cite{Hertzberg21}.
Poor targeting of qubits is a primary cause of crosstalk induced by microwave-frequency drives~\cite{Krinner22}, and can limit gate speed\cite{Magesan20, Ware19, Tripathi19}.
It also increases residual $ZZ$ coupling in processors with always-on qubit-qubit coupling~\cite{Marques22, Krinner22, Chen22}, making gate fidelity and leakage dependent on the state of spectator qubits~\cite{Krinner20}.
For these reasons, qubit frequency targeting has received particular focus in recent years, with laser annealing of constituent Josephson junctions becoming the established post-fabrication trimming method~\cite{Muthusubramanian19, Hertzberg21, Zhang22, Kim22}.
Laser annealing allows selective and controlled reduction of transmon qubit frequencies over a few hundred $\MHz$ without intrinsic impact on coherence times.

Comparatively, post-fabrication trimming of resonators has received less focus to date.
Generally,  CPW resonators for readout and qubit-qubit coupling do not use Josephson junctions and their frequencies are mainly set by geometry~\cite{Pozar05, Goppl08}.
However, unaccounted capacitive loadings and variations in CPW phase velocity $\vphase$ can affect the frequency separation of resonators meant to readout different qubits by frequency-division multiplexing with a common feedline~\cite{Jerger12}. They also prevent the intentional matching of readout resonators with the Purcell-filter resonators used in a recent approach~\cite{Heinsoo18} to break the traditional tradeoff between readout speed and qubit relaxation through the Purcell effect~\cite{Houck08}.
One path to alleviate these problems is making resonators flux tuneable~\cite{Sandberg09}. However, it requires extra on-chip elements such as Josephson-junction loops and dedicated flux-control lines, and can limit the dynamic range.

\begin{figure}[ht!]
\centering
\includegraphics{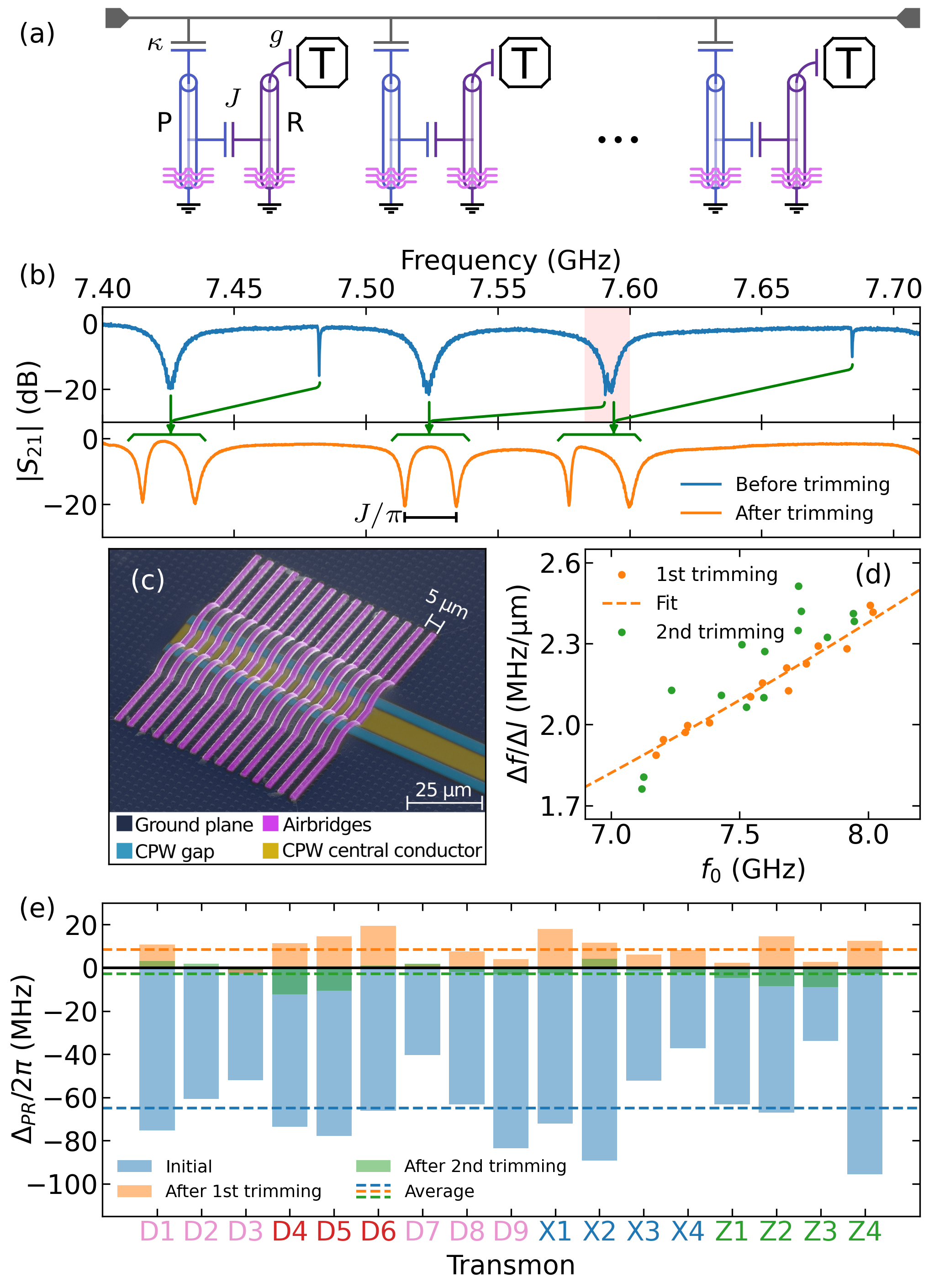}
\caption{
(a) Schematic of the architecture~\cite{Heinsoo18} used to readout multiple transmons (T) using frequency-division multiplexing on a common feedline.
Each transmon has a dedicated readout (R) and Purcell-filter (P) resonator (see text for details).
(b) Top panel: Initial characterization of feedline transmission in the frequency range of three readout-Purcell resonator pairs. The pairs are poorly matched in frequency, as evidenced by the difference in linewidth of their hybridized modes. The red region shows the crowding of readout modes for different transmons.
Bottom panel: Characterization of feedline transmission after one trimming cycle. The matching of R and P resonators is significantly improved and the frequency crowding resolved.
(c) False-colored scanning-electron micrograph of the shoelaces placed near the short-circuit end of a $\lambdaovfour$ CPW resonator.
(c) Change in resonator frequency normalized to the change in length as a function of the pre-trimming frequency. Data are shown for two trimming cycles. The best fit of Eq.~\ref{eq:delta_f_SH} to the data from the first cycle gives $\vphase=1.076\times10^8~\mps$. (d) Detuning $\DeltaPR/2\pi$ between the R and P resonators for all transmons in the Surface-17 processor, measured post-fabrication (blue) and after each of the trimming cycles (orange and green). The dashed lines indicate the average for each characterization.
\label{fig:Reso_trimming}
}
\end{figure}

In this Letter, we introduce a simple airbridge-based method, nicknamed \textit{shoelacing}, enabling frequency trimming of microwave CPW resonators after fabrication and initial characterization. In the circuit QED context, we show that the $100~\MHz$ trimming range with $10~\MHz$ resolution allows correcting frequency mistargeting due to chip design and fabrication uncertainty in 17-transmon circuit QED processors~\cite{Versluis17} (named Surface-17). First, we fix frequency crowding of the resonators used for dispersive readout of different transmons using a common feedline. Next, we demonstrate the frequency matching of dedicated readout and Purcell-filter resonator pairs. Used in combination with transmon trimming by laser annealing, we achieve fast ($400~\ns$) and high-fidelity ($98.6\%$ average) readout on all transmons that are not limited by known extraneous factors. A key advantage of shoelacing is the simultaneous fabrication with conventional CPW airbridges and crossovers, avoiding extra device processing.

Individual transmon readout in Surface-17 is based on the quantum-hardware architecture set forth by Heinsoo \textit{et al.}\cite{Heinsoo18}, illustrated in Fig.~\ref{fig:Reso_trimming}(a). A transmon (T), with qubit transition frequency $\fQ$, couples with strength $g$ to a dedicated $\lambdaovfour$ readout resonator (R) with frequency $\fR$. Their large detuning $\DeltaRQ = 2\pi \times (\fR-\fQ) \gg g/$ makes R shift $\dispshift$ when T is excited from its ground to first-excited state ($\ket{0}$ and $\ket{1}$, respectively). R also couples with strength $\couplingJ$ to a dedicated $\lambdaovfour$ Purcell-filter resonator (P) at $\fP$. Finally, P couples with rate $\kappaP$ to a $50~\Ohm$ feedline whose output connects to an amplification chain with a traveling-wave parametric amplifier (TWPA)~\cite{Macklin15} at its front end.  Ideally,  $\DeltaPR/2\pi = \fP-\fR=0$, so that R and P fully hybridize into two readout modes that frequency-split by $2\couplingJ$, each with effective linewidth $\kappaeff=\kappaP/2$ and dispersive shift $\dispshifteff=\chi$.
High $\kappaeff$ is necessary for fast readout as it sets the rate at which the readout mode builds/depletes internal photon population when the pulse is turned on/off.
Independent readout of multiple transmons using a common feedline is possible by frequency-division multiplexing provided that the hybridized modes for different transmons do not overlap in frequency.
For each transmon, a readout pulse of duration $\taup$ is applied to the feedline input, at a frequency $\drivefreq$ where transmission is dependent on qubit state.

The description above sets the stage for understanding the complications arising when R and P resonators are not well targeted due to chip design error and/or fabrication variability. The example in the top panel of Fig.~\ref{fig:Reso_trimming}(b)
shows two key problems.
First, there is a striking difference in the linewidths of the two modes for each resonator pair, indicating that R and P are poorly matched.
These linewidths are given by
\begin{equation}
    \label{eq:keff_heinsoo}
    \kappaeff = \frac{1}{2}\left(\kappaP \pm \mathrm{Re}\left\{\sqrt{-16\couplingJ^2 + (\kappaP - 2i\DeltaPR)^2}\right\}\right),
\end{equation}
with negative sign for the mode that is more R than P.
Second, there is frequency crowding between modes for different transmons, as highlighted by the shaded red region.

To solve these problems, we place 10 grounding superconducting airbridges ($2~\um$ width and $5~\um$ pitch) at the short-circuited end of each R and P resonator [Fig.~\ref{fig:Reso_trimming}(c)]. Each shoelace contacts the CPW center conductor to the flanking ground planes on both sides, shortening the resonator while preserving the symmetry of the termination.
The removal of the $n$ shoelaces farthest away from the short-circuit on the base layer increases the resonator length by $\Deltal=n\times 5~\um$ and decreases its frequency by
\begin{equation}
    \label{eq:delta_f_SH}
    \Deltaf \approx -\frac{4\freqinitial^2}{\vphase} \Deltal,
\end{equation}
where $\freqinitial$ is the resonator frequency during characterization. The geometry, number and pitch of shoelaces are chosen to enable a $100~\MHz$ trimming range with $10~\MHz$ resolution, and low chance of human error when plucking shoelaces using a fine needle under an optical microscope. While removal allows only a monotonic decrease in resonator frequency, having shoelaces on both R and P resonators allows bidirectional trimming of $\DeltaPR$.

To determine the required trimming $\Deltaf$ for a resonator, an initial characterization is conducted with all transmons biased at their sweetspot (to set the Lamb shift).
Except for cases with evident frequency crowding, $\fR$ and $\fP$ can be determined by fitting a feedline transmission ($\Strans$) measurement to~\cite{Heinsoo18}
\begin{equation}
    \label{eq:S21_heinsoo}
    \Strans(\drivefreq) = 1-\frac{i\kappaP\DeltaR}{4\couplingJ^2 + (2i\DeltaP+\kappaP)2i\DeltaR},
\end{equation}
\noindent
where $\Delta_{\mathrm{P,R}} / 2\pi = f_{\mathrm{P,R}}-\drivefreq$.
Here, $\fR$ includes the Lamb shift induced by transmon coupling~\cite{Koch07}.

One cycle of shoelace plucking solves both the frequency crowding and the poor hybridization of R and P resonators for each transmon, as shown by the bottom panel of Fig.~\ref{fig:Reso_trimming}(b).
An overshoot in $\Deltaf$ is generally observed for this cycle [Fig.~\ref{fig:Reso_trimming}(e)]. This is most likely due to naively approximating Eq.~\ref{eq:delta_f_SH} as $\Deltaf \approx a \Deltal$ with $a = -2~\MHz/\um$. To correct this, we perform a second trimming cycle using Eq.~\ref{eq:delta_f_SH} with $\vphase = 1.076\times10^8~\mps$ extracted from best fits of this equation to first-cycle data. Evidently, the second cycle brings $\DeltaPR$ even closer to target.

\begin{figure}
\centering
\includegraphics{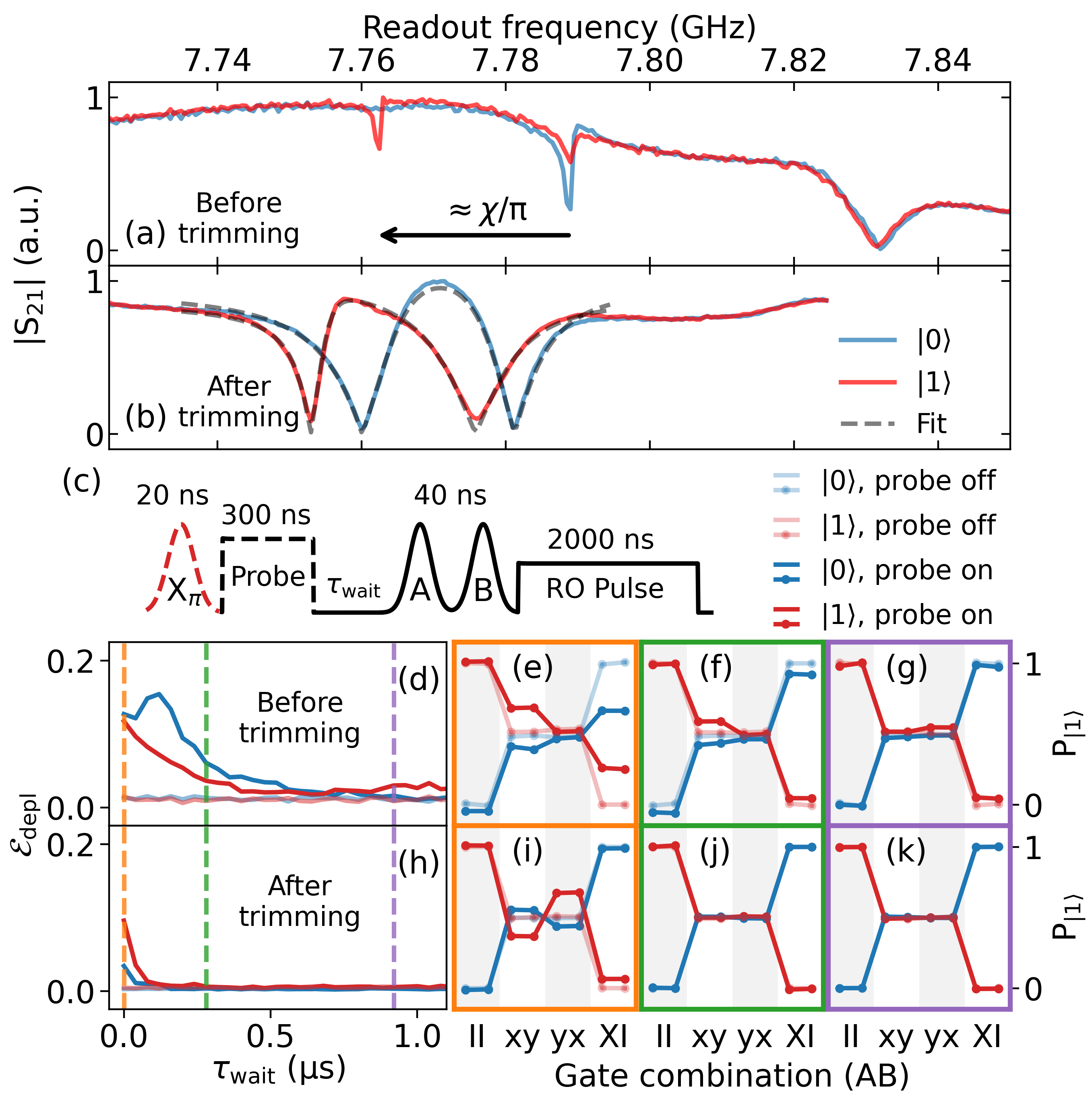}
\caption{
(a) Feedline transmission in the frequency range of the readout-Purcell resonators of transmon $D_4$, when prepared in $\ket{0}$ (blue) and $\ket{1}$ (red). (b) Similar feedline transmission after resonator and qubit frequency trimming. (c) Experiment used to determine the photon depletion time. A probe readout pulse with a fixed duration and amplitude is followed by two back-to-back single-qubit gates $\mathrm{AB} \in \{\mathrm{II, xy, yx, XI}\}$ and a calibrated readout pulse. (d) Deviation from ideal performance of gates AB before trimming. The probe pulse frequency is set to $7.789~\GHz$ ($7.762~\GHz$) for preparation in $\ket{0}$ ($\ket{1}$). The light blue and red traces were acquired without the probe pulse and are displayed for reference. (e-g) Raw measurement outcomes for $\tauwait = 0$, $0.28$ and $0.92~\us$ for the dataset acquired before trimming. (h) Deviation from ideal performance of gates AB, after trimming. The probe pulse frequency is set to $7.760~\GHz$ ($7.753~\GHz$) for preparation in $\ket{0}$ ($\ket{1}$). (i-k) Raw measurement outcomes at the same $\tauwait$ settings as in panels e-g.
\label{fig:Disp_shift}
}
\end{figure}

Matching R and P resonators for high $\kappaeff$ is one of two key ingredients for fast and high-fidelity dispersive readout. The second is matching $|\dispshifteff|=\kappaeff$ to maximize the signal-to-noise ratio at fixed photon number in the readout mode~\cite{Gambetta08}. Transmon frequency trimming by laser annealing can be relied on to achieve this condition because $\fQ$ affects $\dispshift$. In the example of Fig.~\ref{fig:Disp_shift}(a), a low and positive $\DeltaRQ$ caused by a mistargeted high $\fQ$ makes $|\dispshift|$ too large. There is also strong R and P mismatch. In combination, these effects make $|\dispshifteff/\kappaeff|\approx 20$ for the lower-frequency mode (more R than P)  and $\sim 0.01$ for the higher-frequency mode (more P than R). To fix this, we make use of both shoelacing and laser annealing in the same trimming cycle to decrease $\fP$ and $\fQ$, respectively.  The improved matching and decreased $\dispshift$ give $|\dispshifteff/\kappaeff| \approx 0.7$ ($0.4$) for the lower (upper) hybridized mode. Using the mode, the photon depletion time for a fixed readout pulse amplitude decreases from  $>1~\us$ before trimming to $280~\ns$ after trimming, as evaluated with an experiment based on Ref.~\onlinecite{Bultink16} [Fig.~\ref{fig:Disp_shift}(c)].
Optimizing the readout pulse amplitude, $\drivefreq$ and $\taup$ using the procedures presented in the supplement~\cite{SOMtrimming} reaches readout fidelity $\Fro = 98.9\percent$ with a total readout time $\tauro = 400~\ns$ (including $\taup$ and photon depletion time) and keeping the readout highly non-demolition~\cite{Chen22}.
These results conclusively show the benefits of combining resonator and transmon trimming to accelerate and improve qubit readout.

\begin{figure}
\centering
\includegraphics{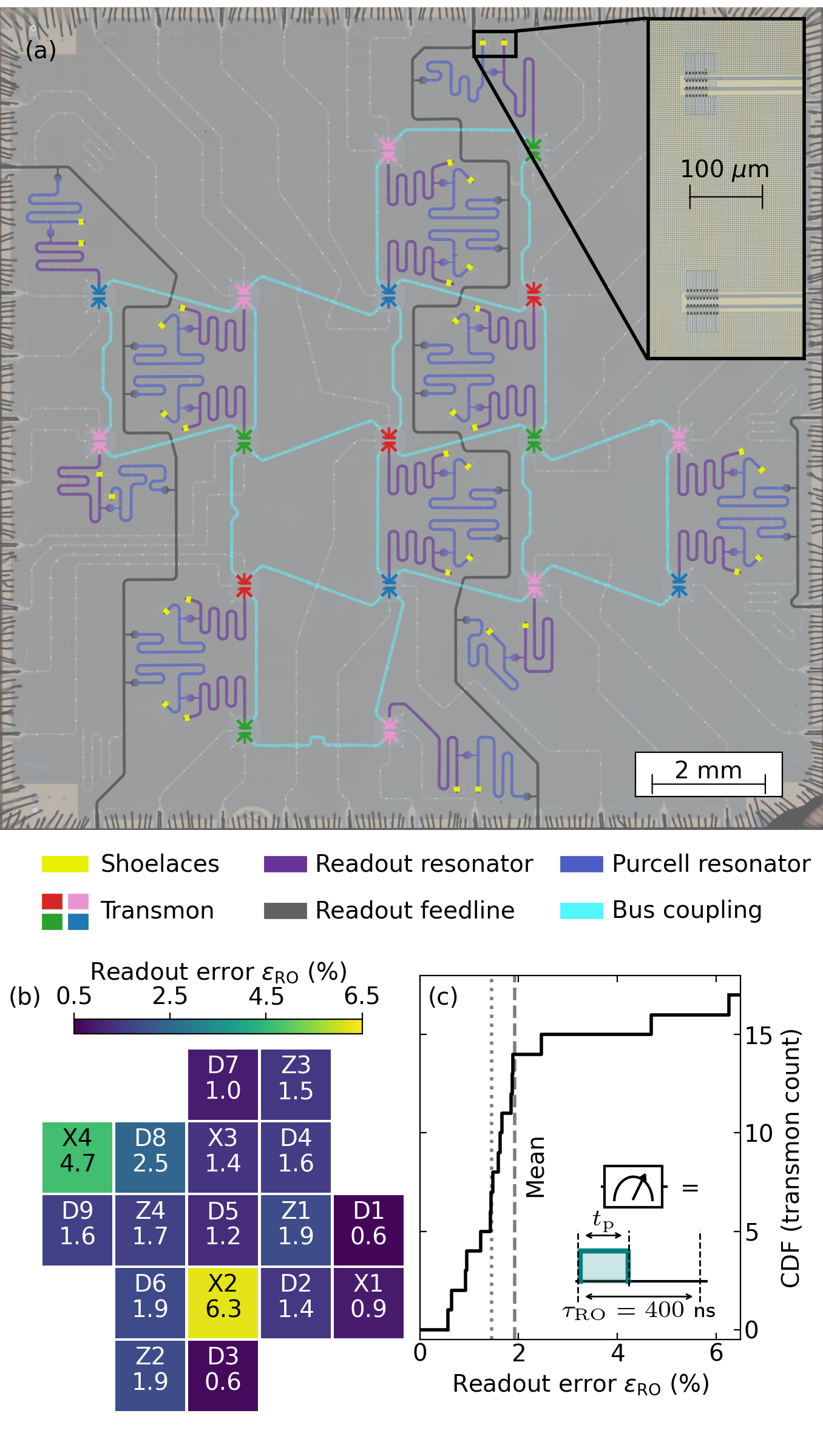}
\caption{
(a) Optical image of our Surface-17 processor used for the distance-3 surface code~\cite{Versluis17}. All transmons have dedicated readout and Purcell resonator pairs with shoelaces enabling their frequency trimming. Inset: Close-up showing the short-circuit ends of one pair. (b) Optimized readout assignment error $\epsro$ for each transmon, with $\tauro=400~\ns$ for all and $\taup$ is optimized for each. (c) Corresponding cumulative distribution function. The mean is $1.92\percent$ when including all transmons (dashed line), and $1.45\percent$ when excluding $X_2$ and $X_4$ (dotted line), whose readout is affected by known extraneous factors (see main text).
\label{fig:S17_readout}
}
\end{figure}

To demonstrate the reliability of combining these trimming methodologies, we entrust them to optimize readout on the actual Surface-17 that we use for QEC experiments. In total, we trimmed 16 of the 34 resonators (either R or P for each case)  and 11 of the 17 transmons.  Figure~\ref{fig:S17_readout}(b) displays the readout assignment error $\epsro=1-\Fro$ achieved for all transmons, with common $\tauro=400~\ns$ and individually optimized $\taup$. The high errors observed in transmons $X_2$ and $X_4$ are due to extraneous sources: $X_2$ is coupled to a two-level system at its sweetspot (the bias point), and feedline transmission at the readout frequency of $X_4$ shows a strong anomalous ripple, attributed to the TWPA. Excluding these, the average $\epsro$ is $1.45\percent$.

In summary, we have introduced and realized shoelacing, a post-fabrication trimming method for CPW resonators. The results showcase the effectiveness of the method in a circuit QED context, solving frequency crowding in readout feedlines and improving the frequency matching of readout and Purcell-filter resonator pairs. Combining shoelacing with transmon frequency trimming by laser annealing reliably achieves fast, high-fidelity readout in multi-transmon processors. We believe that the simplicity and reliability of shoelacing may find uses in microwave-engineering applications beyond circuit QED, including narrow-band matched filters, cameras based on kinetic-inductance detector arrays, and parametric amplifiers, to name a few examples.
\begin{acknowledgments}

We thank A.~Bruno for discussions, and G.~Calusine and W.~Oliver for providing the TWPAs. This research is funded by the 'Quantum Inspire – the Dutch Quantum Computer in the Cloud' project NWA.1292.19.194 of the NWA-ORC program of the Netherlands Organization for Scientific Research (NWO), by Intel Corporation, and by the Office of the Director of National Intelligence (ODNI), Intelligence Advanced Research Projects Activity (IARPA), via the US Army Research Office grant W911NF-16-1-0071. The views and conclusions contained herein are those of the authors and should not be interpreted as necessarily representing the official policies or endorsements, either expressed or implied, of the ODNI, IARPA, or the U.S. Government.

Correspondence and requests for materials should be addressed to L.D.C. (l.dicarlo@tudelft.nl).
The data presented are available at
\verb"http://github.com/DiCarloLab-Delft/"\\
\verb"Post_Fabrication_Trimming_Data".

\end{acknowledgments}

\bibliographystyle{apsrev4-2}

\input{Bibliography/bibliography_main}
\end{bibunit}
\newpage

\renewcommand{\theequation}{S\arabic{equation}}
\renewcommand{\thefigure}{S\arabic{figure}}
\renewcommand{\thetable}{S\arabic{table}}
\renewcommand{\bibnumfmt}[1]{[S#1]}
\renewcommand{\citenumfont}[1]{S#1}
\setcounter{figure}{0}
\setcounter{equation}{0}
\setcounter{table}{0}

\begin{bibunit}[apsrev4-2]

\onecolumngrid
\section*{Supplementary material for `'\nametitle''}
\FloatBarrier

\subsection{Fabrication details}

Device fabrication begins by sputtering of a $\NbTiN$ layer (target thickness $200~\nm$) on a $\Silicon$ substrate followed by etching to define all circuit elements except transmon Josephson junctions, shoelaces and conventional CPW airbridges and crossovers. Transmon junctions are fabricated next, and shoelaces, airbridges and crossovers are conveniently fabricated last using common processing steps. For this last step, a layer of PMGI SF15 is initially spun at $2500~\rpm$ and patterned using a wide electron beam at $500~\uCcm$ dose. Next, this first resist layer is developed by dipping the device in a solution of AZ400K and water (1:5 for $60~\seconds$ and 1:10 for $30~\seconds$) before a $30~\seconds$ water rinse. The rounded structure of the airbridges is obtained by reflowing the resist at $225~\Celsius$ for $5~\minutes$. Two more layers of PMMA A8 495K and PMMA 950K A8 are then spun at $1000~\rpm$ and $2000~\rpm$, respectively, and patterned using a $38~\nA$ beam at a dose of $1100~\uCcm$, before development with a 1:2 solution of MIBK and IPA. A RIE O2 plasma cleaning at $20~\Watt$ for $45~\seconds$ is then applied before evaporation, followed by a $30~\seconds$ dip in 1:7 BOE to remove any residual oxides. Finally, a Plassys FC-2000 system is used to evaporate $100~\nm$ Al + $200~\nm$ Ti + $100~\nm$ Al before lifting off the resist bilayer by dipping the device in anisole for $12~\hours$.

\subsection{Resonator pair transmission response}

Following the theory of Ref.~\onlinecite{Heinsoo18}, feedline transmission near a readout-Purcell resonator pair in our system is given by
\[
\Strans= 1 - \frac{\kappaP(\gammaR + 2i \DeltaR + \kappaDrive)}{4\couplingJ^2 + (\gammaP + 2i \DeltaP + \kappaP)(\gammaR + 2i \DeltaR + \kappaDrive)},
\]
where $\kappaDrive$ is the decay rate of the readout resonator through the transmon drive line, and  $\gammaR$ and $\gammaP$ are the intrinsic loss rates of the readout and Purcell resonators, respectively.
Assuming $\kappaDrive, \gammaP, \gammaR \approx 0$ gives Eq.~\ref{eq:S21_heinsoo}.

\subsection{Study of resonator intrinsic quality factor}

Losses in the chosen readout mode set an upper bound to the fraction of readout photons that contribute to the detected readout signal, given by
$\Qi/2\left(\Qc+\Qi\right)$, where $\Qc$ and $\Qi$ are the mode effective coupling and intrinsic quality factors, respectively.
The factor of 2 is due to having an open feedline (i.e., without the input capacitor of Ref.~\onlinecite{Heinsoo18}).
Note that when readout and Purcell resonators are properly frequency matched in our processors, $\Qc\sim10^3$ for the readout modes.

\begin{figure}
\centering
\includegraphics{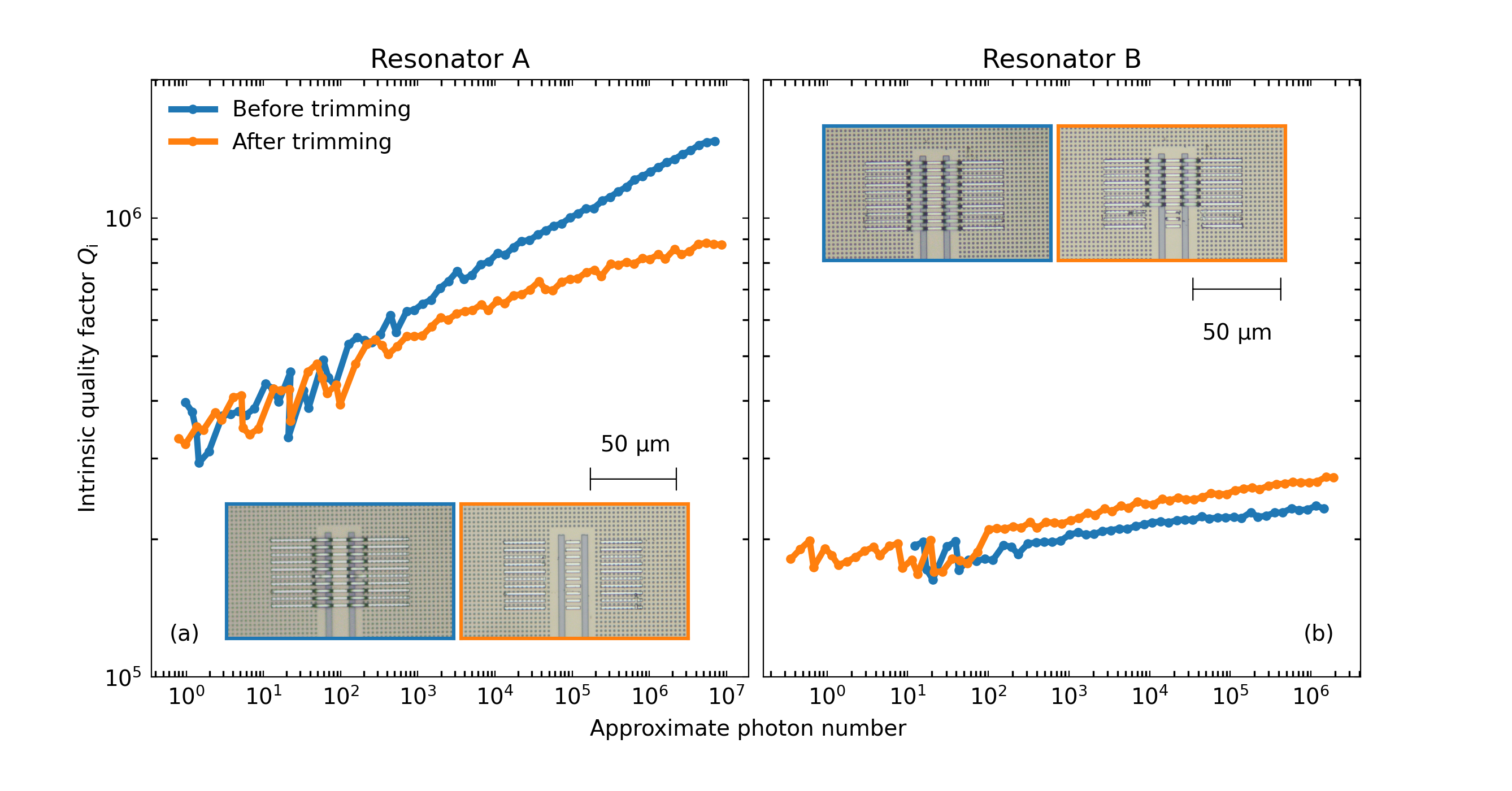}
\caption{
Comparison of the intrinsic quality factor $\Qi$ as a function of approximate photon number, before (blue) and after (orange) total (a) and partial (b) removal of shoelaces in two test resonators with coupling quality factor $\Qc \approx 10^5$. Insets: optical images of the short-circuited end of the test resonators, before and after shoelace removal.
\label{fig:Quality_factor}
}
\end{figure}

Figure~\ref{fig:Quality_factor} studies $\Qi$ for $\lambdaovfour$ resonators (with shoelaces) coupled to a common feedline in a test device.
The $\Qi$ was extracted as a function of approximate intra-resonator photon number using the frequency-domain methods described in Ref.~\onlinecite{Bruno15}.
To increase the sensitivity of $\Qi$ extraction in these test resonators, we chose $\Qc\sim10^5$.
We observe $\Qi\sim10^5$ in the quantum regime (few-photon level), both before and after partial or total shoelace removal.
This convincingly shows that any internal loss contributed by shoelaces in our readout-Purcell resonator pairs is insignificant in terms of its impact on the quantum efficiency of the readout chain.

\subsection{Transmon frequency trimming}

The qubit transition frequency $\fQ$ and anharmonicity of a transmon (always biased to the sweetspot in our case) are measured using standard spectroscopy methods~\cite{Krantz19}. Using these, a numerical inversion procedure reliably extracts the transmon charging energy $\Ec$ and the maximum Josephson coupling energy $\Ej$. By the Ambegaokar-Baratoff relation~\cite{Ambegaokar63}, $\Ej \propto \resJ^{-1}$, where $\resJ$ is the normal-state resistance of the pair of Josephson junctions (in parallel) forming the transmon SQUID loop. The sweetspot $\fQ$, approximately $\left( \sqrt{8\Ej \Ec}-\Ec \right) / h$, can be trimmed post-fabrication using laser annealing~\cite{Hertzberg21, Zhang22, Kim22}, which is proven to selectively and controllably increase $\resJ$.

\begin{figure}
\centering
\includegraphics[width=0.95\columnwidth]{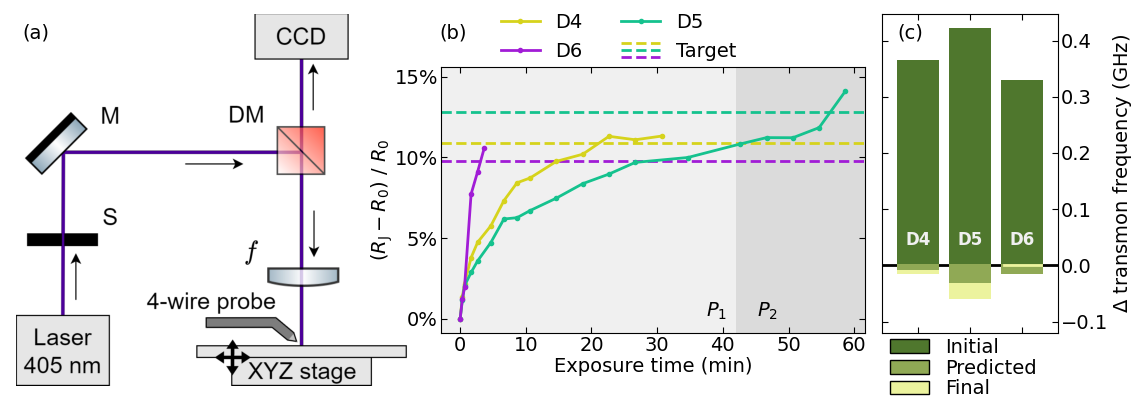}
\caption{
Trimming of transmon qubit frequency by  closed-loop laser annealing.
(a) Schematic of our homebuilt laser annealing setup.
The exposure time of a  diode laser is controlled using a beam shutter (S).
The beam is routed through a dichroic mirror (DM) which acts as a beamsplitter and wavelength cutoff filter.
An objective lens (f) focuses the beam on the device plane.
The reflected light passes back through the objective lens and dichroic mirror into the camera (CCD) which is used for visual feedback during alignment.
(b) Change in 4-wire resistance $\resJ$ for three transmons in the device of Fig.~\ref{fig:Disp_shift}, normalized to the initial resistance $R_0$, as a function of incident laser power and exposure time.
Horizontal lines indicate the predetermined target resistance for each transmon. The two regions correspond to incident laser powers $P_1=170~\mW$ and $P_2=200~\mW$.
(c) Bar plot of transmon qubit frequency deviation from target at three key points: initial cryogenic characterization, prediction based on measured $\resJ$ after annealing, and final cryogenic characterization.
\label{fig:LASIQ}
}
\end{figure}

Our homebuilt setup for automated laser annealing and resistance measurement [Fig.~\ref{fig:LASIQ}(a)] contains a diode laser with $405~\nm$ wavelength and $50~\um$ diameter beam spot, sufficient to uniformly cover both junctions forming the SQUID loop of one transmon. A 4-wire measurement probes $\resJ$ free of any contact resistance between the probe needles and the transmon capacitive plates.
To allow a closed-loop annealing cycle, the 4-wire probe is motor-controlled in the $Z$ direction (i.e., out of plane), while the relative sample position is controlled using a motorized XY stage.
This control allows alternating between measuring $\resJ$ and aligning for laser exposure.
The normalized change in 4-wire resistance is tracked as a function of incident laser power and exposure time for each transmon throughout the automated annealing procedure [Fig.~\ref{fig:LASIQ}(b)].
The cycle-exit criteria are evaluated individually for each transmon and are met once either $\resJ$ reaches a certain target,
or the expected next-cycle exposure time exceeds a threshold. If the first criterion is satisfied, the transmon is considered to be successfully trimmed.
Meeting the second criteria interrupts the annealing procedure and requires an increase in incident laser power.
For the example in Fig.~\ref{fig:LASIQ}(b), showing annealing results for transmons $D_4$-$D_6$ in the device of Fig.~\ref{fig:Disp_shift}, increasing the power from $170~\mW$ to $200~\mW$ is needed to increase the tunability range in $D_5$. A power increase reduces the expected exposure time per fractional resistance change. Figure~\ref{fig:LASIQ}(c) compares the initial (measured before annealing), predicted (from extracted $\Ec$ and $\resJ$ post-annealing), and actual final (measured post-annealing) sweetspot qubit frequencies. Deltas are with respect to the targeted final qubit frequency for each transmon.

\subsection{Readout optimization and characterization}

\begin{figure}
\centering
\includegraphics[width=0.7\columnwidth]{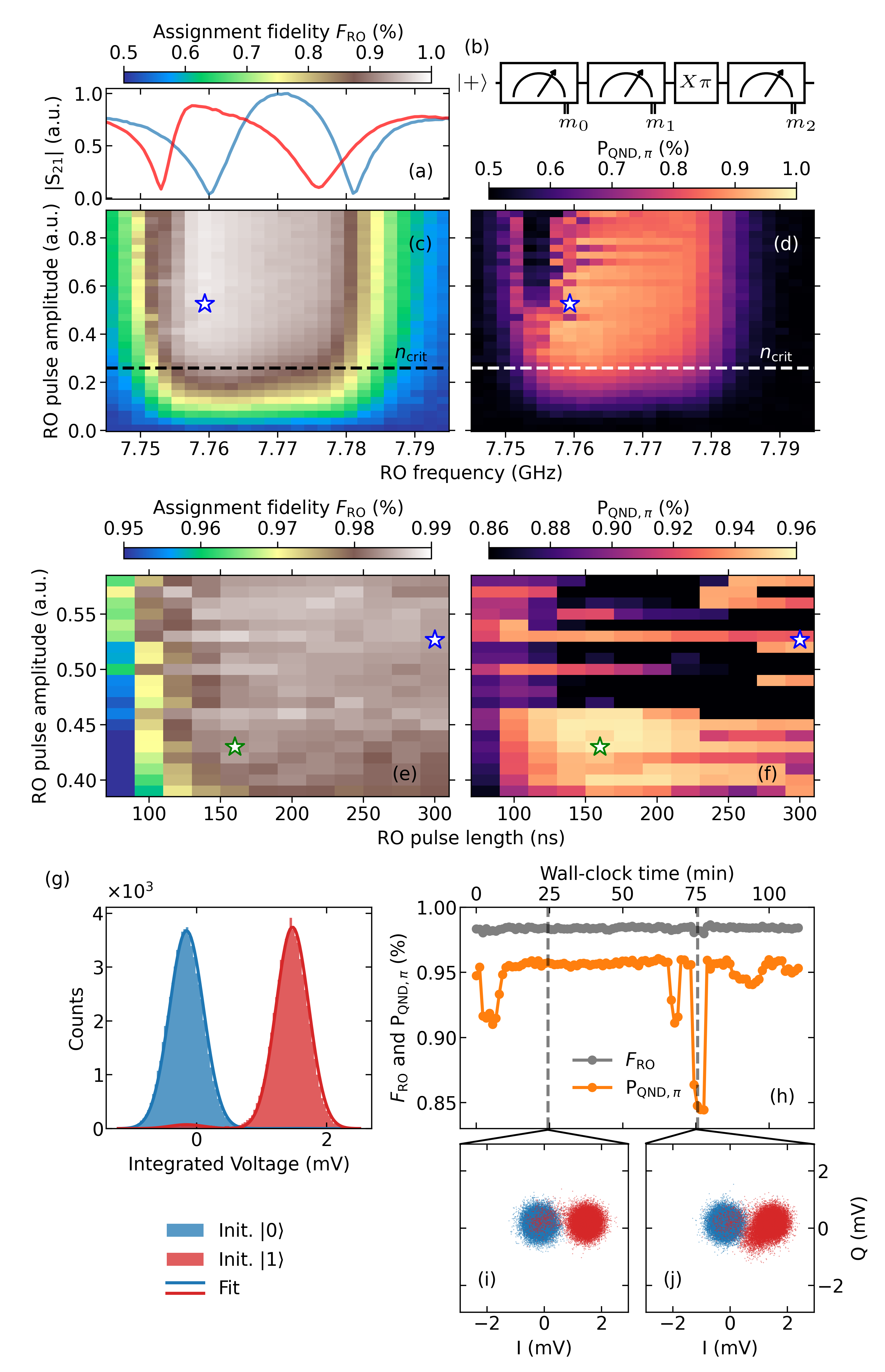}
\caption{
(a) Same feedline transmission measurements as in Fig.~\ref{fig:Disp_shift}(a), for reference. (b) Pulse sequence used to determine $\pqndpi$, following Ref.~\onlinecite{Chen22} (c, d) $\Fro$ and $\pqndpi$ as a function of readout pulse amplitude and frequency, at constant $\taup = 300~\ns$. The blue star indicates the maximum $\pqndpi$. The critical photon number is estimated for the readout mode at $7.760~\GHz$. (e,f) $\Fro$ and $\pqndpi$ as a function of $\taup$ and amplitude at a fixed $\drivefreq = 7.760~\GHz$ and  $\tauro = \taup + 280~\ns$. The blue star indicates the starting point from the calibration in panels (c) and (d), while the green star shows the point of maximum $\pqndpi$, at $\taup = 180~\ns$. (g) Single-shot readout ($\SSRO$) histograms at the optimal point, yielding $\Fro = 98.9\percent$. (h) Stability analysis of $\Fro$ and $\pqndpi$ measured over two hours. (i,j) Raw data of $\SSRO$ shots in the IQ plane during at different times. Panel (j) shows that measurement induces leakage from $\ket{1}$ to $\ket{2}$ during intermittent periods during which $\pqndpi$ decreases by up to $\sim10\percent$.
\label{fig:Readout_calib}
}
\end{figure}

Here we describe the typical procedure used to optimize transmon readout and benchmark its performance.
We use transmon $D_4$ (same device and transmon as in Fig.~\ref{fig:Disp_shift}).
These benchmarks are the average assignment fidelity $\Fro$ and the quantum non-demolition probability $\pqndpi$ defined in Ref.~\onlinecite{Chen22}:
\[
\pqndpi = \left[p(m_1=0|m_2=1) + p(m_1=1|m_2=0)\right] / 2,
\]
where $m_1$ and $m_2$ are the outcomes of the last two measurements in Fig.~\ref{fig:Readout_calib}(b). We first measure these two benchmarks as a function of readout amplitude and frequency, with constant $\taup = 300~\ns$. We typically observe that the best readout performance is achieved slightly beyond the linear dispersive linear, a factor $\sim4$ above the estimated critical photon number $\ncrit$ [Figs.~\ref{fig:Readout_calib}(c,d)]. The relation between $\ncrit$ and the readout pulse amplitude is calibrated by measuring the AC Stark shift imposed on the transmon by a continuous readout tone~\cite{Schuster05}. We choose the combination of readout amplitude and frequency that yields the highest $\pqndpi$ as a first estimate. A subsequent sweep of the readout amplitude and $\taup$ at fixed frequency indicates that a better performance can be achieved with a shorter pulse [Figs.~\ref{fig:Readout_calib}(e,f)]. The new set of optimal parameters (green cross in Figs.~\ref{fig:Readout_calib}(e,f)) gives $\Fro=98.9\percent$ [Fig.~\ref{fig:Readout_calib}(g)] assuming no initialization errors, which we minimize as much as possible using post-selection on an initial measurement  indicating transmon in $\ket{0}$. Repeated measurements extracting $\Fro$ and $\pqndpi$ over 2 hours [Fig.~\ref{fig:Readout_calib}(h-j)] indicate that these optimal readout conditions are not completely stable. We observe short, intermittent periods during which $\pqndpi$ decreases by up to $\sim10\percent$. Plots of single shots in the IQ plane clearly show that the measurement excites leakage from $\ket{1}$ to $\ket{2}$ during these bursts. We do not fully understand the source of this instability.

\subsection{Device information}
The data shown are taken from four devices: three Surface-17s and one test device. Data for each main-text figure are taken from a different Surface-17. Data in Figs.~\ref{fig:LASIQ} and ~\ref{fig:Readout_calib} are from the Surface-17 of Fig.~\ref{fig:Disp_shift}. Data in Fig.~\ref{fig:Quality_factor} are taken from the test device, which contains only resonators (with shoelaces) coupling to one common feedline.

\bibliographystyle{apsrev4-2}

\input{Bibliography/bibliography_som}
\end{bibunit}

\end{document}

%% file: Preamble/packages.tex
\usepackage[makeroom]{cancel}
\usepackage{dcolumn}
\usepackage{makecell}
\usepackage[utf8]{inputenc}
\usepackage[T1]{fontenc}
\usepackage{mathptmx}
\usepackage{commath}
\usepackage{graphicx}
\usepackage{multirow}
\usepackage{amsmath}

\usepackage{soul}
\usepackage{verbatim}
\usepackage{bm}
\usepackage{outlines}
\usepackage{siunitx}
\usepackage[space]{grffile}
\usepackage{upgreek}
\usepackage[dvipsnames]{xcolor}
\usepackage{xr}
\usepackage{color}
\usepackage{tikz}
\usepackage{mathtools}
\usepackage{amssymb,placeins}
\usepackage[globalcitecopy]{bibunits}
\usetikzlibrary{quantikz}

\newcommand\myshade{85}
\colorlet{mylinkcolor}{violet}
\colorlet{mycitecolor}{YellowOrange}
\colorlet{myurlcolor}{Aquamarine}
\usepackage{hyperref}
\hypersetup{
  linkcolor  = mylinkcolor!\myshade!black,
  citecolor  = mycitecolor!\myshade!black,
  urlcolor   = myurlcolor!\myshade!black,
  colorlinks = true,
}
\usepackage[capitalize]{cleveref}
\usetikzlibrary{calc, positioning} 